# CHAPTER 15 – ASSESSING COMPETENCY USING METACOGNITION AND MOTIVATION: THE ROLE OF TIME-AWARENESS IN PREPARATION FOR FUTURE LEARNING

**Mark Abdelshiheed, Mehak Maniktala, Tiffany Barnes, and Min Chi**
North Carolina State University

## Introduction

One fundamental goal of learning is being prepared for future learning (PFL) (Bransford & Schwartz, 1999) and able to extend acquired skills and problem-solving strategies to different domains and environments. While substantial research has shown that PFL can be accelerated by obtaining metacognitive skills (Zepeda et al., 2015; Chi & VanLehn, 2010) or influenced by the individual's motivation (Belenky & Nokes-Malach, 2013; Nokes & Belenky, 2011), no prior work investigated whether the interaction of the two factors could assess students' competency for PFL. In this chapter, we tackle this research question in one type of highly interactive e-learning environment, Intelligent Tutoring Systems (ITSs). More specifically, we investigate whether the combination of metacognitive skills and motivation would assess students' learning abilities in logic, and their competence to extend these abilities to a subsequent domain, probability.

We focus on two types of metacognitive skills related to problem-solving strategies: strategy-awareness and time-awareness, that is respectively, how and when to use each strategy. Moreover, we track the accuracy collected from the online traces from both ITSs to measure students' motivation levels. By doing so, we hypothesize that high-motivated students who additionally know how and when to use each strategy will yield the highest learning outcomes on a logic tutor, and transfer their acquired skills to a subsequent probability tutor. Both tutors have been extensively evaluated in the past decade and a series of papers have been published to show their effectiveness independently (Chi & VanLehn, 2010; Barnes et al., 2008; Chi & Vanlehn, 2007).

In deductive task domains such as logic and probability, solving a problem often requires producing a proof, argument, or derivation consisting of one or more inference steps, and each step is the result of applying a domain principle, rule, or operator. Prior work has shown that students often use a mixture of problem-solving strategies such as forward chaining (FC) and backward chaining (BC) during their problem solving (Priest & Lindsay, 1992; Simon & Simon, 1978; Newell & Simon, 1972). Many prior studies investigated the impact of teaching students an explicit problem-solving strategy on their learning gains (Zepeda et al., 2015; Chi & VanLehn, 2007) or compared students who were taught different types of strategies (Boden et al., 2018; Chi & VanLehn, 2010). In this chapter, we argue that time-awareness should be considered as an independent type of metacognitive skills apart from problem-solving strategies, and we investigate:

1. how students' knowledge about how to use a problem-solving strategy (strategy-awareness) and when to use it (time-awareness) would impact their learning
2. how such impact would change when we consider the student motivation
3. how would the interactions between the two types of metacognitive skills and motivation impact students' learning in a new domain, probability.

## Background

Bransford and Schwartz (1999) proposed the theory of Preparation for Future Learning (PFL) that states that students need to continue to learn, and investigates whether they are prepared to do so. Similar to prior work (Chi & VanLehn, 2010), we bring PFL into the ITS context, where it is possible to directly observe behaviors associated with PFL. In this chapter, we evaluate students' choices of how and when to select a problem-solving strategy. Based on Winne and Azevedo (2014), mastering how to use each strategy is a cognitive skill, but when incorporated with awareness about when such strategy should be used, it becomes a metacognitive skill. Therefore, we consider strategy-awareness and time-awareness to be two different types of metacognitive skills. Specifically, we investigate how the interactions of the two types of metacognitive skills and motivation could impact PFL.

### The Impact of Metacognitive Skills on PFL

Metacognition indicates one's realization of their self-cognition as well as being able to regulate and understand it (Chambres et al., 2002; Roberts & Erdos, 1993). It is the act of exercising and monitoring control of cognitive skills (Efklides, 2011). Hence, we consider that a metacognitive skill consists of a cognitive skill and a regulator for controlling this skill. Many studies have shown that metacognitive skills have positive impacts on learning (Zepeda et al., 2019), on students' learning behaviors (Belenky & Nokes, 2009), and on PFL across ITSs (Zepeda et al., 2015; Chi & VanLehn, 2010). Several approaches have been used to evaluate metacognitive skills, such as strategy selection (Chi & VanLehn, 2010; Roberts & Erdos, 1993), tutoring prompts (Zepeda et al., 2015; Belenky & Nokes, 2009), and reading comprehension and memory recall (Chambres et al., 2002).

Zepeda et al. (2015) demonstrated that metacognitive instruction could influence student metacognitive skills, motivation, and transfer learning. Students who were taught planning, monitoring, and evaluating, made better metacognitive judgments and showed higher motivation levels (e.g. self-efficacy and task value) than those who were not taught these skills. As an example of PFL, the former also performed better on a novel self-guided learning task than the latter.

In our prior work, Chi and VanLehn (2010) investigated the transfer of metacognitive skills from a probability tutor to a physics tutor. We showed that an ITS teaching domain-independent metacognitive skills could close the gap between high and low learners, not only in the domain where they were taught (probability), but also in a second domain where they were not taught (physics). In that study, the metacognitive skills included a problem-solving strategy and principle-emphasis instructions. We found that it was the principle-emphasis skill that is transferred across the two domains and that closed the gap between the high and low learners. In this chapter, we investigate how students' own metacognitive skills (strategy- and time-awareness) would impact their learning and also prepare them for future learning in a new domain with a new ITS.

### The Impact of Motivation on PFL

Substantial work has shown the impact of motivation on PFL (Belenky & Nokes-Malach, 2013, 2012; Nokes & Belenky, 2011). For instance, Belenky and Nokes-Malach (2012) found that PFL is influenced by the interaction of students' motivation and different forms of instruction. They found that students who had high mastery-approach goal orientation showed signs of transfer, irrespective of the instruction type. Furthermore, students who were allowed to innovate new strategies developed higher motivation aspects, compared to those who followed direct instruction. Later, the same innovative students showed strong evidence of PFL when given a final wording problem. Nokes and Belenky (2011) incorporated students' achievement goals into a PFL framework that accounts for transfer success or failure. The framework represents a loop of goal generation, environment interpretation, knowledge & goal representation, solution generation, and solution evaluation. The last step decides whether the loop will be repeated or not. After testing this framework, they reported that mastery-approach goal-oriented students were more likely to succeed in knowledge transfer.

One of the crucial questions for research on motivation is how to define and measure motivation. Eccles (1983) defined motivation to be the individual's perception of three factors: expectations for success, subjective task value, and intrinsic interest. The three factors respectively represent how much success one would expect from a task, what value it has and why would there be an interest to accomplish it. Touré-Tillery and Fishbach (2014) classified motivation into two dimensions: process-focused 'doing it right' and outcome-focused 'getting it done'. To measure motivation, prior research explored self-efficacy (Kalender et al., 2019; Boden et al., 2018; Zepeda et al., 2015), goal orientation (Otieno et al., 2013; Belenky & Nokes-Malach, 2013, 2012), and accuracy (Touré‑Tillery & Fishbach, 2014). The majority of these studies used surveys to measure these aspects. For instance, Kalender et al. (2019) used a survey to measure three motivational aspects based on the achievement goals: self-efficacy, interest, and sense of belonging. In recent years, digital technologies such as ITSs made it possible to measure motivation using students' online traces in ITS (Otieno et al., 2013). Otieno et al. (2013) used the online use of hint and glossaries in a geometry tutor as a motivation measure and found that the online measures differ from the motivation measures using questionnaire data, and the former was more predictive of posttest scores than the latter. Therefore, in this chapter, we use students' online traces to measure their motivation levels. More specifically, we extract the initial accuracy from each tutor to measure students' motivation.

## Experiment

Our data were collected from an undergraduate computer science course at North Carolina State University across three semesters. Students were trained on the logic tutor first, and then on the probability tutor six weeks later. The tutors were assigned as one of their regular homework assignments and the completion of both tutors was required for full credit. Students were told that the assignment would be graded based on the demonstrated effort, not performance. A total of 495 students finished both tutors with the following distribution: $N = 151$ for Fall 2017, $N = 128$ for Spring 2018, and $N = 216$ for Fall 2018.

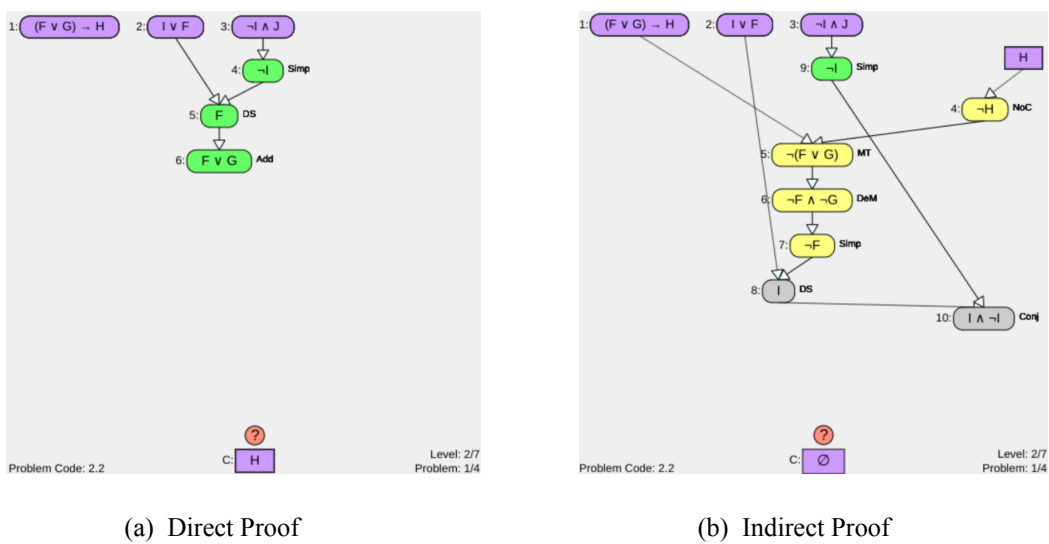

(a) Direct Proof          (b) Indirect Proof

**Figure 1: Logic Tutor Problem-Solving Strategies**

## Methods

### The Logic and Probability tutors

Students went through a standard pretest-training-posttest procedure on each tutor. The logic tutor teaches students propositional logic proofs. A student can solve a problem in one of two strategies: direct or indirect. Figure 1a shows that for direct proofs, a student needs to derive the conclusion node at the bottom from the givens at the top; while Figure 1b shows that for indirect proofs, a student derives a contradiction from the givens and the negation of the conclusion. Both logic pre- and post-test have two problems and their scores are a function of time and accuracy. The training on the logic tutor includes 20 problems.

Figure 2 shows the graphical user interface (GUI) for the probability tutor that teaches students how to apply principles to solve probability problems. The pre- and post-test sections have 14 and 20 problems, respectively. These problems require students to derive an answer by writing and solving one or more equations. The training includes 12 problems.

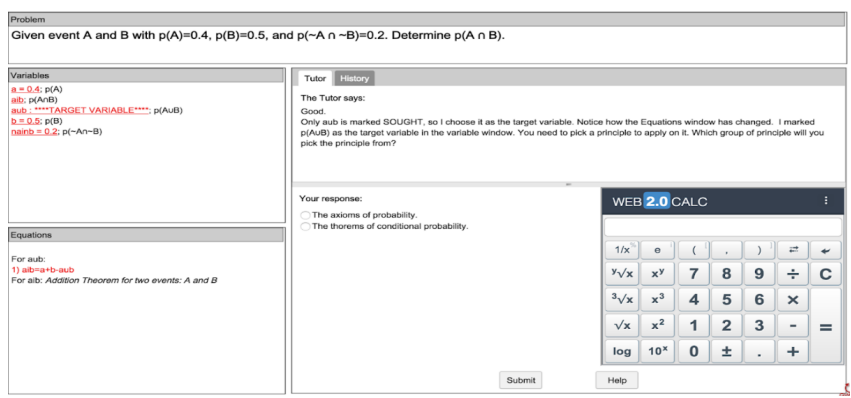

**Figure 2: Probability Tutor Interface**

There are two major differences between the two ITSs:

1. In the probability tutor, the pre- and post-test scores are based on accuracy. Both tests were graded in a double-blind manner by experienced graders using a partial-credit rubric. In the logic tutor, they are based on both accuracy and efficiency. Since there are only two questions in each test, the class instructor believes that it is as important for students to solve them accurately as for them to solve them quickly. For comparison purposes, all test scores are normalized to the range of [0, 100]. Note that in both tutors, the posttest is much harder than the pretest.
2. In the logic tutor, students can select FC-like direct proofs (the default), or choose to switch to BC-like indirect proofs. Conversely, in the probability tutor, students can only use BC during training. In both tutors, students can employ any strategy during the pre- and post-test.

### Metacognitive Skills

Students can choose to switch problem-solving strategies only when training on the logic tutor. Hence, we measure students' metacognitive skills based on their interactions with the logic tutor alone. The training section in the logic tutor is organized into five ordered levels with an incremental degree of difficulty and each level consists of four problems. Each problem can be solved by either following the default FC (direct) or switching to BC (indirect). However, most advanced problems (in higher levels) can be solved much more efficiently by BC. Therefore, we expect that effective problem solvers should switch their strategy on these problems, and more importantly, they should switch it early when solving them. Thus, our metacognitive skill measurement is a combination of strategy-awareness: using the default direct proof or switching to indirect proof (Chi & VanLehn, 2010; Chambres et al., 2002; Roberts & Erdos, 1993), and time-awareness: when such switch happens (Winne & Azevedo, 2014). We consider two factors in time-awareness: one is that a student should switch in later levels (harder training problems) where the savings will be big and the other is that students should switch early (when convenient) during solving a problem. On average, students take 210 actions to solve a problem, and the median number of actions that a student takes before switching is 30. Therefore, we calculated the metacognitive score (MetaScore) for a student **i** as:

$$MetaScore_i = \sum_{L=1}^{5} \left[ \sum_{p=1}^{4} [L * SAware_{ip} * TAware_{ip}] \right]$$

where strategy-awareness $SAware_{ip} = 1$ indicates that student i switched strategy when solving problem p at level L, while 0 means no switch. For time-awareness, $TAware_{ip} = 1$ indicates that the student i switched early on problem p ($\leq$ 30 actions) while $TAware_{ip} = -1$ is for a late switch (> 30 actions). As stated before, the training levels have an incremental degree of difficulty, as each level introduces a new logic rule. Since the rate of change of rules per level is constant, the difficulty of the tutor was assumed to be linear, and therefore, we weighted the strategy- and time-awareness by the corresponding level number. Based on this formula, $MetaScore_i > 0$ suggests that student i is both strategy-aware and time-aware; if $MetaScore_i < 0$, it shows that student i is strategy-aware but not good at knowing when to switch (time-unaware). Finally, if $MetaScore_i = 0$, we do not have enough evidence on the student's metacognitive skills in that he may simply follow the default FC settings. Based on MetaScores, students are classified into three groups: those who show both strategy- and time-awareness (MetaScore > 0) are referred to as the '**Str_Time**' group ($N = 145$); those who showed strategy awareness only (MetaScore < 0) as '**Str_Only**' ($N = 166$); and the default students (MetaScore = 0) as '**Default**' ($N = 184$).

*Motivation*

Inspired by prior research (Touré‑Tillery & Fishbach, 2014), we measured students' motivation by tracking the accuracy of their online traces. By doing so, we acknowledge that students often have different motivations: some are more process-focused for learning the domain as much as possible and some are more outcome-focused for better grades. Similar to prior work (Vollmeyer & Rheinberg, 2006; Rheinberg et al., 2000), we define students' motivation based on their initial interactions in the early stages of each tutor. More specifically, we use the percentage of correct rule applications in the first two problem-solving questions as our motivation indicators. In other words, our measured students' initial motivation levels do not consider that students' motivation levels may change over time. Students are divided into high- and low-motivation groups through a median split. For Logic: $HM_{Logic}(N = 248)$ and $LM_{Logic}(N = 247)$, and for Probability: $HM_{Prob}(N = 249)$ and $LM_{Prob}(N = 246)$. A chi-square test showed no significant evidence on students staying in the same motivation level from the logic tutor to the probability one: $\chi^2(1, N = 495) = 1.26, p = .26$. In other words, students' motivation levels may change over a course of a semester or change due to the subjective domains. Additionally, our motivation definition differs from students' incoming competence in that one-way ANOVA showed no significant difference in the pretest scores between the high- and low-motivation students: $F(1, 493) = 0.7, p = .17$ for Logic and $F(1, 493) = 0.001, p = .98$ for Probability.

## Results

We will examine the impact of 1) metacognitive skills alone, 2) motivation alone, and 3) the interactions of the two on students' learning across both tutors. For each tutor, students' learning performance is measured using their corresponding pre- and post-test scores, together with their normalized learning gain (NLG) defined as: $(NLG = \frac{Post - Pre}{\sqrt{100 - Pre}})$ (Abdelshiheed, Hostetter, Shabrina, et al., 2022; Abdelshiheed, Hostetter, Yang, et al., 2022; Abdelshiheed et al., 2021, 2020), where 100 is the maximum posttest score. For reporting convenience, we normalize the pre and post scores to the range of [0, 100].

Table 1: Comparing the three Metacognitive Groups in each tutor

| Group | Size | Logic Tutor | | | Probability Tutor | | |
|---|---|---|---|---|---|---|---|
| | | Pre | Post | NLG | Pre | Post | NLG |
| Str_Time | 145 | 78.4 (3.2) | 75.8 (1.7) | 0.94 (.395) | 72.3 (2.8) | 75.5 (3) | 0.02 (.06) |
| Str_Only | 166 | 74.9 (3) | 68.2 (1.67) | -0.46 (.39) | 72.1 (2.5) | 74 (2.8) | 0.01 (.05) |
| Default | 184 | 75.5 (2.8) | 70.9 (1.68) | 0.19 (.393) | 71.8 (2.6) | 73.4 (2.6) | -0.007 (.05) |

**Metacognitive Skills**

Table 1 above compares the three metacognitive groups' learning performances on the logic and probability tutors. It shows the mean and SD of the pretest scores (Pre), the posttest scores (Post), and the NLGs. For the logic tutor, while we found no significant difference among the three groups on Pre, a one-way ANCOVA analysis with the metacognitive group as a factor and the pretest score as a covariate showed a significant difference in their posttest scores: $F(2, 491) = 17.3, p < .001, \eta = 0.3$. Subsequent contrast analyses showed that Str_Time scored significantly higher than both Str_Only: $t(309) = 5.8, p < .0001, d = 4.5$ and Default: $t(327) = 3.8, p < .001, d = 2.9$. Additionally, Default scored significantly higher than Str_Only: $t(348) = 2.2, p = .03, d = 1.6$. For NLG, while a one-way ANOVA showed no significant difference among the three groups on the logic NLG, subsequent contrast analyses showed that Str_Time scored significantly higher than Str_Only: $t(309) = 2.4, p = .02, d = 3.6$. For the probability tutor, however, no significant results were found among the three metacognitive groups on either Pre, Post, or NLG.

To summarize, these results suggest that strategy-awareness alone cannot lead students to learn better in logic; students need to be time-aware as well. Additionally, while Str_Time group learned significantly better than Str_Only and Default in logic, this was not observed in probability. For Str_Only students, it seems they were negatively affected by their lack of time-awareness, given the aforementioned note that the posttest is much harder than the pretest.

**Motivation Level**

Table 2 compares the learning performance of the high- and low-motivation groups on the logic and probability tutors. As mentioned before, no significant difference was found between the high- and low-motivation groups on the pretest on either tutor. As expected, a one-way ANCOVA analysis using motivation as a factor and pretest as a covariate showed that on both tutors, high-motivation students scored significantly higher than their low peers on the corresponding posttest: $F(1, 492) = 15.8$, $p < .001$, $\eta = 0.17$ for logic and $F(1, 492) = 24.5$, $p < .001$, $\eta = 0.17$ for probability. For the NLG, While no significant difference was found between the two groups' logic NLG, one-way ANOVA showed that highly motivated students had significantly higher probability NLG than their low peers: $F(1, 493) = 7.6$, $p < .01$, $\eta = 0.12$. In short, this suggests that our motivation measure is reasonable in that: the highly motivated students indeed significantly outperformed their low-motivated peers on the posttest on both the logic and probability tutors. They also had significantly higher NLG on the probability tutor.

Table 2: Comparing the Motivation Level in each tutor

| Logic Tutor | | | | |
|---|---|---|---|---|
| Group | Size | Pre | Post | NLG |
| $HM_{Logic}$ | 248 | 78.9 (5.3) | 73.6 (1.4) | 0.25 (.06) |
| $LM_{Logic}$ | 247 | 73.4 (5.5) | 69.2 (1.4) | 0.14 (.07) |
| Probability Tutor | | | | |
| Group | Size | Pre | Post | NLG |
| $HM_{Prob}$ | 249 | 81.7 (4.2) | 79 (1.8) | 0.05 (.04) |
| $LM_{Prob}$ | 246 | 77 (4.4) | 69 (2.5) | -0.03 (.04) |

## Interaction Between Metacognition and Motivation

**Logic Tutor**

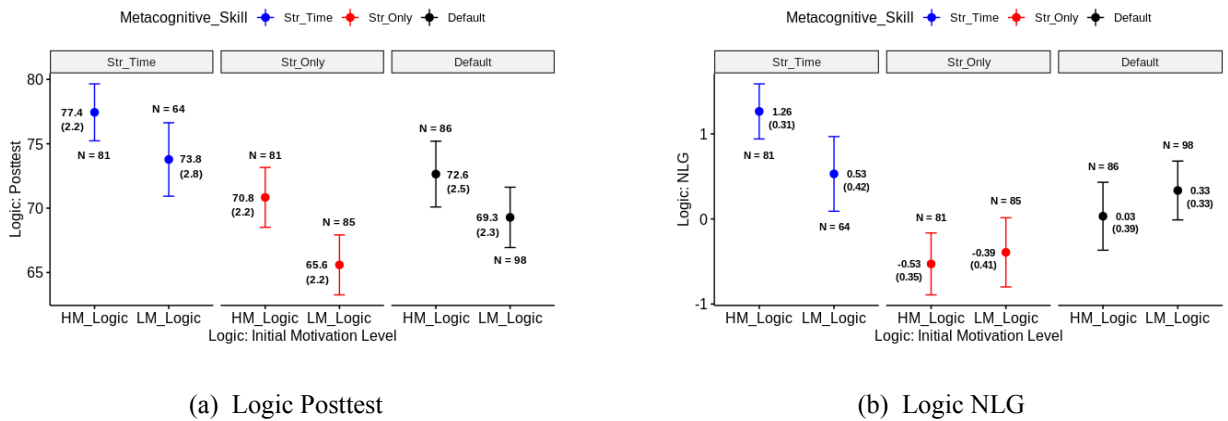

(a) Logic Posttest    (b) Logic NLG

**Figure 3: Logic Performance: Metacognition and Motivation**

Combining the three metacognitive groups (Str_Time, Str_Only, and Default) with the two motivation levels ($HM_{Logic}$ and $LM_{Logic}$) resulted in six groups. A chi-square test showed no significant difference in the distribution of motivation level across the metacognitive groups: $\chi^2(2, N = 495) = 2.87$, $p = .24$. Additionally, no significant difference was found among the six groups on the logic pretest: $F(2, 489) = 0.69, p = .49$.

Figure 3 compares the six groups' performance on the logic tutor. Regarding the logic posttest (**Fig. 3a**), a two-way ANCOVA using the metacognitive groups and motivation levels as factors and pretest as a covariate showed no significant interaction effect. However, there was a main effect of metacognitive groups: $F(2, 488) = 16.6, p < .0001$, and a main effect of motivation level: $F(1, 488) = 16.7, p < .0001$. More specifically, within each metacognitive group, $HM_{Logic}$ significantly outperformed the corresponding $LM_{Logic}$ group: $t(143) = 2, p = .04, d = 1.4$ for Str_Time, $t(164) = 3.1, p < .01, d = 2.4$ for Str_Only and $t(182) = 2.1, p = .03, d = 1.4$ for Default. Among the three $HM_{Logic}$ groups, the high-motivation Str_Time students scored significantly higher than their peers: $t(160) = 3.8, p < .001, d = 3$ against the high-motivation Str_Only peers and $t(165) = 2.8, p < .01, d = 2.1$ against the high-motivation Default ones.

For the NLG (**Fig. 3b**), a two-way ANOVA using the same two factors found no significant interaction effect nor any main effect. However, among the three $HM_{Logic}$ groups, the high-motivation Str_Time scored significantly higher than both high-motivation Str_Only: $t(160) = 2.3, p = .03, d = 5.4$ and high-motivation Default: $t(165) = 2.2, p = .03, d = 3.5$. No significant difference was found among the three $LM_{Logic}$ groups. Additionally, only within the two Str_Time groups, the high-motivation students scored significantly higher than their low-motivation peers. No significant difference between the high- and low-motivation groups was found within Default and Str_Only. In short, our results suggest that the high-motivation Str_Time group performs the best among the six groups in terms of both posttest and NLG scores on the logic tutor.

**Probability Tutor**

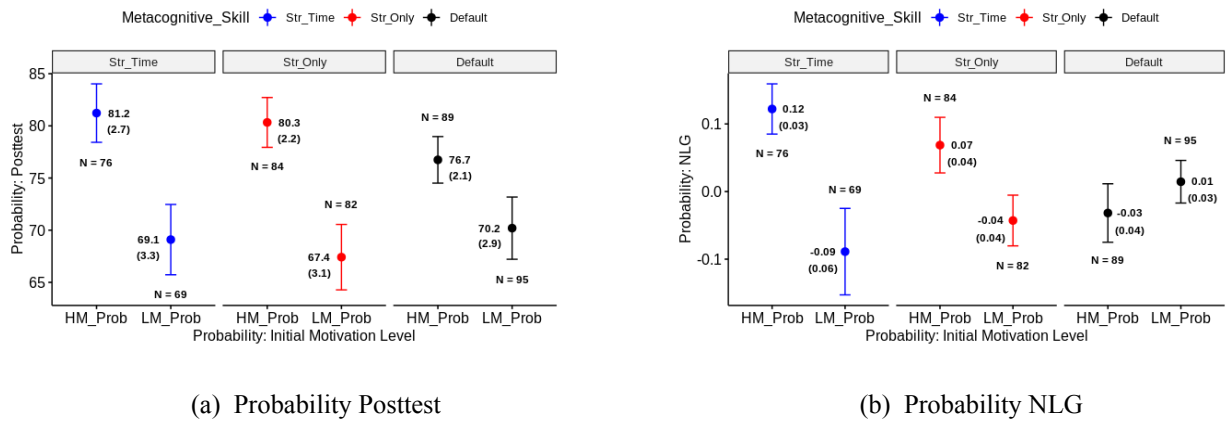

(a) Probability Posttest

(b) Probability NLG

**Figure 4: Probability Performance: Metacognition and Motivation**

Similarly, we combined the three metacognitive groups with the two motivation levels defined based on students' interactions on the probability tutor: $HM_{Prob}$ and $LM_{Prob}$, resulting in six groups. A chi-square test showed students' motivation level on the probability tutor did not differ significantly across the three metacognitive groups: $\chi^2(2, N = 495) = 0.53, p = .76$. Moreover, we found no significant difference between the six groups on the probability pretest: $F(2, 489) = 0.5, p = .63$.

Figure 4 depicts the performance of the six groups on the probability tutor. For the posttest (**Fig. 4a**), a two-way ANCOVA using metacognitive skills and motivation as factors and pretest scores as a covariate showed a significant interaction effect: $F(2, 488) = 3.8, p = .02, \eta = 0.09$. Additionally, there was a main effect of motivation in that high-motivation students scored significantly higher than their low peers: $F(1, 488) = 24.4, p < .0001$. Among the three highly motivated groups, both Str_Time and Str_Only scored significantly higher than Default: $t(163) = 2.4, p = .02, d = 1.9$ and $t(171) = 2.4, p = .02, d = 1.7$, respectively. However, no such difference was found among the three low-motivation groups.

Similarly, as shown in **Figure 4b**, a two-way ANOVA using metacognitive skills and motivation as factors showed a significant interaction effect on the probability NLG: $F(2, 489) = 6.4, p < .01, \eta = 0.16$ and there was also a main effect of motivation: $F(1, 489) = 7.8, p < .01$. Subsequent contrast analyses showed that high-motivation Str_Time students scored significantly higher than their low peers: $t(143) = 3.8, p < .001, d = 4.4$. The same pattern was observed between the two Str_Only groups: $t(164) = 2.2, p = .03, d = 2.9$. Across the three high-motivation groups, both Str_Time and Str_Only scored significantly higher than their Default peers: $t(163) = 3, p < .01, d = 4.2$ and $t(171) = 2, p = .04, d = 2.5$, respectively. In short, on our probability tutor, the high-motivation Str_Time group performs the best among the six groups, on both posttest scores and NLGs.

## Conclusions and Discussions

In this chapter, we investigate how two factors, metacognitive skills and motivation, would impact student learning across two domains: logic and then probability. Our results from analyzing 495 students' performance on two tutors show that when considering each factor alone, no consistent robust pattern is found. However, when we combine the two factors, we find that students who are highly motivated, strategy-, and time-aware consistently outperform their peers across both domains.

Firstly and most importantly, our analyses confirm the importance of motivation in that across both tutors, the impacts of metacognitive skills on student learning are only observed among the highly motivated student groups. For low motivated students, no significant difference was found among the three metacognitive groups in either tutor. In other words, our results reveal an aptitude-treatment interaction (ATI) effect (Kanfer & Ackerman, 1989) in that some students may be insensitive to learning unless the presented material matches their aptitude. While such findings are not surprising, they suggest that it is crucial to further understand why certain students lack motivation, and to explore how to motivate them. Moreover, our findings indicate that our choice of using students' online accuracy traces on the first two questions is a reasonable way to measure their motivation levels.

Secondly, while problem-solving strategies have been extensively explored in prior research, as far as we know this is the first work that investigates students' metacognitive skills from both strategy-aware and time-aware aspects. Our results suggest that these two skills are indeed different in that while both Str_Time and Str_Only groups know about problem-solving strategies, only the former knows when to apply them. More importantly, it is essential to consider the time-aware aspect when assessing students' metacognitive skills in that when highly motivated, Str_Time consistently outperforms their Str_Only and Default peers on both tutors.

Thirdly, our results show that Str_Only can benefit greatly by training on an ITS that explicitly teaches and follows problem-solving strategies. While the high-motivation Str_Only performed the worst than their high-motivation peers on the logic tutor, they performed as well as the high-motivation Str_Time and both outperformed their Default peers on the probability tutor. One potential explanation is that the time-aware aspect of the skills is not needed when training on the probability tutor, since it follows the same explicit problem-solving strategy on all problems.

Finally, we emphasize the importance of mastering different problem-solving strategies for highly motivated students, and its role on PFL. We found that only across the highly motivated groups, both Str_Only and Str_Time had significantly higher probability scores than the Default group. This finding suggests evidence for metacognitive skill transfer for highly motivated students who are also aware of switching strategies. To sum up, while time awareness could be a decisive factor for consistency, strategy awareness might identify students who are prepared for future learning.

Despite these findings, it is important to note that there are at least two caveats in our analyses. First, we measured students' motivation using the first two problems on each tutor and we did not consider that students' motivation levels may vary during the training. Also, the probability tutor supports only one problem-solving strategy. A more convincing testbed would be to use any ITS that supports different types of strategies, so we can investigate whether students can properly use them.

## Recommendations and Future Research

This chapter reinforces the significance of understanding how and when to apply each problem-solving strategy. This is consistent with prior findings that within multi-strategy domains, it is insufficient to only learn what each strategy is. Rather, it is equally important to learn when to use each. Therefore, Generalized Intelligent Framework for Tutoring (GIFT) can benefit from this by prioritizing the importance of mastering **how** and **when** to use each strategy. For example, GIFT can ensure that individuals understand the methodology, context, timing, and reasoning beyond a given strategy.

For future work, we will investigate whether explicitly teaching different problem-solving strategies would boost the performance of students who lack strategy- or time-awareness so that they can catch up with their peers. Additionally, we will explore different ways of motivating students and meeting their expectations about the content and interface of the tutors.